\RequirePackage{fix-cm}
\documentclass[twocolumn,compsoc]{CVM}

\graphicspath{{figures/},{figures/cvm/}}

\def\ManuscriptInfo{%Manuscript received: 2014-12-31; accepted: 2015-01-30.
}

\def\PaperTitle{A Survey on Deep Geometry Learning: From a Representation Perspective}

\def\AuthorNames{Yun-Peng Xiao$^1$, Yu-Kun Lai$^2$, Fang-Lue Zhang$^3$, Chunpeng Li$^1$, Lin Gao$^1$(\rm{\Letter})}

\def\AuthorAdress{
    $1\quad $ & Institute of Computing Technology, Chinese Academy of Sciences, Beijing, China. E-mail: \href{mailto:xiaoypgk@gmail.com}{xiaoypgk@gmail.com}, \href{mailto:cpli@ict.ac.cn}{cpli@ict.ac.cn}, \href{mailto:gaolin@ict.ac.cn}{gaolin@ict.ac.cn }(\rm{\Letter}) \\
    $2\quad $ & School of Computer Science and Informatics, Cardiff University, Wales, UK. E-mail: \href{mailto:LaiY4@cardiff.ac.uk}{LaiY4@cardiff.ac.uk} \\
    $3\quad $ & School of Engineering and Computer Science, Victoria University of Wellington, New Zealand. E-mail: \href{mailto:fanglue.zhang@ecs.vuw.ac.nz}{fanglue.zhang@ecs.vuw.ac.nz} \\
}

\def\firstheader{DOI 10.1007/s41095-xxx-xxxx-x\hfill Vol. x, No. x, month year, xx--xx\\[5mm]~Review Article}

\headevenname{{Xiao \etal} }

\begin{document}

\MakePageStyle

\MakeAbstract{Researchers have now achieved great success on dealing with 2D images using deep learning. In recent years, 3D computer vision and Geometry Deep Learning gain more and more attention. Many advanced techniques for 3D shapes have been proposed for different applications. Unlike 2D images, which can be uniformly represented by regular grids of pixels, 3D shapes 
have various representations, such as depth and multi-view images, voxel-based representation, point-based representation, mesh-based representation, implicit surface representation, etc. However, the performance for different applications largely depends on the representation used, and there is no unique representation that works well for all applications. 
Therefore, in this survey, we review recent development in deep learning for 3D geometry from a representation perspective, 
summarizing the advantages and disadvantages of different representations in different applications. We also present existing datasets in these representations and further discuss future research directions. 
}

\MakeKeywords{3D representation, geometry learning, neural networks, computer graphics}

\section{Introduction}\label{sec:introduction}
Recent improvements in methods for acquisition and rendering of 3D models haven resulted in consolidated repositories containing massive amounts of 3D shapes on the Internet. With the increased availability of 3D models, we have been seeing an explosion in the demands of %acquiring 
processing, generation and visualization of 3D models in a variety of disciplines, such as medicine, architecture and entertainment. The techniques for matching, identification and manipulation of 3D shapes have become fundamental building blocks in modern computer vision and computer graphics systems. Due to the complexity and irregularity of 3D shape data, how to effectively represent 3D shapes remains a challenging problem. Thus, there have been extensive research efforts concentrating on how to deal with and generate 3D shapes based on different representations.

In early research on 3D shape representations, 3D objects were normally modeled with a global approach, such as constructive solid geometry and deformed superquadrics. Those approaches have several drawbacks when utilized for the tasks like recognition and retrieval. First, when representing imperfect 3D shapes, including those with noise and incompleteness, which are common in practice, such representations may 
 impose negative influence on matching performance. Second, the high-dimensionality heavily burdens the computation and tends to make models overfit. Hence, more sophisticated methods are designed to extract representations of 3D shapes in a more concise, yet discriminative and informative form. % in a compact space.

Several related surveys have been published~\cite{bronstein2017geometric, ahmed2018deep, guo2019deep}, which focus on different aspects of deep learning for 3D geometry. Moreover, with rapid development of 3D shape representations and related techniques for deep learning, it is essential to further summarize up-to-date research works.
In this survey, we mainly review deep learning methods on 3D shape representations and discuss their advantages and disadvantages considering different application scenarios. We now give a brief summary of different 3D shape representation categories.

%The quick development of the computer vision in 2D Images field benefits from the specific representation, pixels. The unified, regular and efficient data representation makes it easy for neural network to extract features and generate new images. However, when it comes to 3D shapes, things become more complex. There is more than one representations for 3D shapes. The neural network layer operations and architectures which perform well in 2D images may not suitable for 3D fields. Roughly, the 3D shape representations include depth and multi-view images, voxel-based representation, surface representation, implicit surface representation, structured representation and deformation representation.

\textbf{Depth and multi-view images} can be used to represent 3D models in the 2D field. The regular structure of images makes them efficient to be processed. Depending on whether depth maps are included, 3D shapes can be presented by RGB (color) or RGB-D (color and depth) images viewed from different viewpoints. Because of the influx of available depth data due to the popularity of 2.5D sensors, such as Microsoft Kinect, Intel RealSense, etc., multi-view RGB-D images are widely used to represent real-world 3D shapes. The large asset of image-based processing models can be leveraged using this representation. But it is inevitable that this kind of representation loses some geometry features.

A \textbf{voxel} is a 3D extension of the concept of pixel. Similar with pixels in 2D, the voxel-based representation also has a regular structure in the 3D space. The architectures of some neural networks which have been demonstrated useful in the 2D image  field~\cite{krizhevsky2012imagenet, lecun2010convolutional} can be easily extended to the voxel form. Nevertheless, adding one dimension means an exponentially increased data size. As the resolution increases, the amount of required memory and computational costs increase dramatically, which restricts the representation only to low resolutions when representing 3D shapes.

\textbf{Surface-based representation} describes 3D shapes by encoding their surfaces, which can also be regarded as 2-manifolds. Point clouds and meshes are both discretized forms of 3D shape surfaces. Point clouds use a set of sampled 3D point coordinates to represent the surface. It can be easily generated by scanners but difficult to process due to their lack of order and connectivity information. Researchers use order invariant operators such as the max pooling operator in deep neural networks~\cite{qi2017pointnet, qi2017pointnet++} to mitigate the lack of order problem.
Meshes can depict higher quality 3D shapes with less memory and computational cost compared with point clouds and voxels. A mesh contains a vertex set and an edge set. Due to its graphical nature, researchers have made attempts to build graph-based convolutional neural networks for coping with meshes. Some other methods regard meshes as the discretization of 2-manifolds. Moreover, meshes are more suitable for 3D shape deformation. One can deform a mesh model by transforming vertices while keeping the connectivity at the same time. %
%%%ZFL: is it necessary to mention the paper here?
%\cite{wang2018pixel2mesh} used this property and reconstruct 3D shapes from single view images by deforming a template ellipsoid mesh, but the reconstructed 3D shapes has a fixed topological structure, which limit the generation capacity on mesh-based representation.

\textbf{Implicit surface representation} exploits implicit field functions, such as occupancy functions~\cite{mescheder2019occupancy} and signed distance functions~\cite{xu2019disn}, to describe the surface of 3D shapes. The implicit functions learned by deep neural networks define the spatial relationship between points and surfaces. They provide a description with infinite resolution of 3D shapes with reasonable memory consumption, and are capable of representing shapes with changing topology. Nevertheless, implicit representations cannot reflect the geometric features of 3D shapes directly, and usually need to be transformed to explicit representations such as meshes. Most methods apply iso-surfacing, such as marching cubes~\cite{lorensen1987marching}, which is an expensive operation.

\textbf{Structured representation}. One way to cope with complex 3D shapes is to decompose them into structure and geometric details, leading to structured representations.
Recently, increasingly more methods regard a 3D shape as a collection of parts and organize them linearly or hierarchically. The structure of 3D shapes is processed by \textit{Recurrent Neural Networks (RNNs)}~\cite{zou20173d}, \textit{Recursive Neural Networks (RvNNs)}~\cite{li2017grass} or other network architectures. Each part of the shape can be processed by unstructured models. The structured representation focuses on the relations (such as symmetry, supporting, being supported, etc.) between different parts within a 3D shape, which provides better description capability than alternative representations.

\textbf{Deformation-based representation}.
Unlike rigid man-made 3D shapes such as chairs and tables, there are also a large number of non-rigid (e.g. articulated) 3D shapes such as human bodies, which also play an important role in computer animation, augmented reality, etc. The deformation-based representation is proposed mainly for describing the intrinsic deformation properties while ignoring the extrinsic transformation properties. Many methods use rotation-invariant local features for describing shape deformation to reduce the distortion and keep the geometry details at the same time.

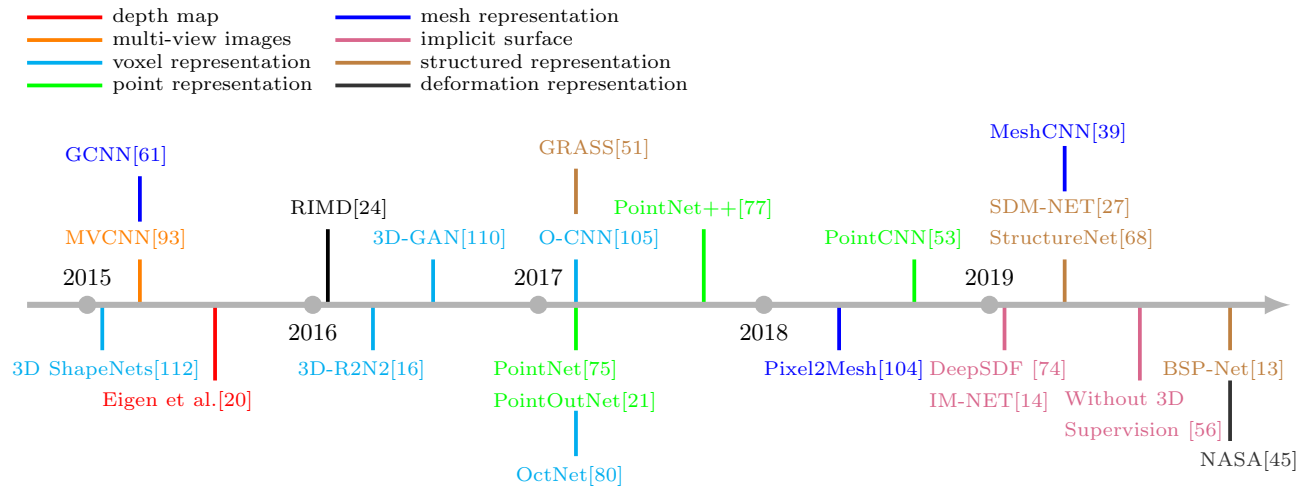
\begin{figure*}[htbp]
\centering
\begin{tikzpicture}
    %\draw[black, thick] (0,0) rectangle (17.2, 8);
    \fill[gray!60] (1,4) circle (1.2mm) node[black, anchor=south, yshift=1.5mm] {2015};
    \fill[gray!60] (4,4) circle (1.2mm) node[black, anchor=north, yshift=-1.5mm] {2016};
    \fill[gray!60] (7,4) circle (1.2mm) node[black, anchor=south, yshift=1.5mm] {2017};
    \fill[gray!60] (10, 4) circle (1.2mm) node[black, anchor=north, yshift=-1.5mm] {2018};
    \fill[gray!60] (13, 4) circle (1.2mm) node[black, anchor=south, yshift=1.5mm] {2019};
    \draw[red, line width=0.5mm] (0.2, 7.8) -- (1.2, 7.8) node[black, anchor=west] {\small{depth map}};
    \draw[orange, line width=0.5mm] (0.2, 7.5) -- (1.2, 7.5) node[black, anchor=west] {\small{multi-view images}};
    \draw[cyan, line width=0.5mm] (0.2, 7.2) -- (1.2, 7.2) node[black, anchor=west] {\small{voxel representation}};
    \draw[green, line width=0.5mm] (0.2, 6.9) -- (1.2, 6.9) node[black, anchor=west] {\small{point representation}};
    \draw[blue, line width=0.5mm] (4.3, 7.8) -- (5.3, 7.8) node[black, anchor=west] {\small{mesh representation}};
    \draw[purple!60, line width=0.5mm] (4.3, 7.5) -- (5.3, 7.5) node[black, anchor=west] {\small{implicit surface}};
    \draw[brown, line width=0.5mm] (4.3, 7.2) -- (5.3, 7.2) node[black, anchor=west] {\small{structured representation}};
    \draw[black!80, line width=0.5mm] (4.3, 6.9) -- (5.3, 6.9) node[black, anchor=west] {\small{deformation representation}};
    \draw[cyan, line width=0.5mm] (1.2, 4) -- (1.2, 3.4) node[anchor=north, text width=3cm, xshift=3mm] {\small{3D ShapeNets\cite{wu20153d}}};
    \draw[orange, line width=0.5mm] (1.7, 4) -- (1.7, 4.6) node[anchor=south, text width=3cm, xshift=0.5cm] {\small{MVCNN\cite{su2015multi}}};
    \draw[blue, line width=0.5mm] (1.7, 5.1) -- (1.7, 5.7) node[anchor=south, text width=2cm] {\small{GCNN\cite{masci2015geodesic}}};
    %\draw[yellow, line width=0.5mm] (2.2, 4) -- (2.2, 5) node[anchor=south, text width=3cm, xshift=1.2cm] {\small{VoxNet\cite{maturana2015voxnet}}};
    \draw[red, line width=0.5mm] (2.7, 4) -- (2.7, 3) node[anchor=north, text width=3cm] {\small{Eigen et al.\cite{eigen2014depth}}};
    \draw[black, line width=0.5mm] (4.2, 4) -- (4.2, 5) node[anchor=south, text width=3cm, xshift=1cm] {\small{RIMD\cite{gao2016efficient}}};
    \draw[cyan, line width=0.5mm] (4.8, 4) -- (4.8, 3.4) node[anchor=north, text width=2cm] {\small{3D-R2N2}\cite{choy20163d}};
    \draw[cyan, line width=0.5mm] (5.6, 4) -- (5.6, 4.6) node[anchor=south, text width=2cm, xshift=0.2cm] {\small{3D-GAN}\cite{wu2016learning}};
    \draw[green, line width=0.5mm] (7.5, 4) -- (7.5, 3.4) node[anchor=north, text width=2cm, xshift=-1mm] {\small{PointNet\cite{qi2017pointnet} PointOutNet\cite{fan2017point}}};
    \draw[cyan, line width=0.5mm] (7.5, 2.6) -- (7.5, 2) node[anchor=north,text width=3cm, xshift=7mm] {\small{OctNet\cite{riegler2017octnet}}};
    \draw[cyan, line width=0.5mm] (7.5, 4) -- (7.5, 4.6) node[anchor=south, text width=3cm, xshift=1cm] {\small{O-CNN\cite{wang2017cnn}}};
    \draw[brown, line width=0.5mm] (7.5, 5.2) -- (7.5, 5.8) node[anchor=south, text width=3cm, xshift=1cm] {\small{GRASS\cite{li2017grass}}};
    \draw[green, line width=0.5mm] (9.2, 4) -- (9.2, 5) node[anchor=south, text width=3cm, xshift=3mm] {\small{PointNet++\cite{qi2017pointnet++}}};
    \draw[blue, line width=0.5mm] (11, 4) -- (11, 3.4) node[anchor=north, text width=3cm, xshift=5mm] {\small{Pixel2Mesh\cite{wang2018pixel2mesh}}};
    \draw[green, line width=0.5mm] (12, 4) -- (12, 4.6) node[anchor=south, text width=3cm, xshift=3mm] {\small{PointCNN\cite{li2018pointcnn}}};
    \draw[purple!60, line width=0.5mm, text width=3cm] (13.2, 4) -- (13.2, 3.4) node[anchor=north, text width=2cm] {\small{DeepSDF \cite{park2019deepsdf} IM-NET\cite{chen2019learning}}};
    \draw[brown, line width=0.5mm] (14, 4) -- (14, 4.6) node[anchor=south, text width=2cm] {\small{SDM-NET\cite{gao2019sdm} StructureNet\cite{mo2019structurenet}}};
    \draw[blue, line width=0.5mm] (14, 5.5) -- (14, 6.1) node[anchor=south, text width=2cm, yshift=-1mm] {\small{MeshCNN\cite{hanocka2019meshcnn}}};
    \draw[purple!60, line width=0.5mm, text width=3cm] (15, 4) -- (15, 3) node[anchor=north, xshift=5mm] {\small{Without 3D Supervision~\cite{liu2019learning}}};
    \draw[brown, line width=0.5mm, text width=3cm] (16.2, 4) -- (16.2, 3.4) node[anchor=north, xshift=0.6cm] {\small{BSP-Net\cite{chen2019bsp}}};
    \draw[black!80, line width=0.5mm, text width=2cm] (16.2, 3) -- (16.2, 2.2) node[anchor=north, xshift=6mm] {\small{NASA\cite{jeruzalski2019nasa}}};
    
    \draw[gray!60, -latex, line width=0.8mm] (0.2, 4) -- (17, 4);
\end{tikzpicture}
\caption{The timeline of deep learning based methods for various 3D shape representations.}
\label{fig:voxeltimeline}
\end{figure*}

Recently, deep learning has achieved superior performance in contrast to classical methods in many fields, including 3D shape analysis, reconstruction, etc. A variety of architectures of deep networks have been designed to process or generate 3D shape representations, which we refer to as \emph{geometry learning}. In the following sections, we focus more on most recent deep learning based methods for representing and processing 3D shapes in different forms. According to how the representation is encoded and stored, our survey is organized in the following structure: Section~\ref{sec:images} reviews image-based shape representation methods. Sections \ref{voxels} and \ref{sec:surfaces} introduce voxel-based and surface-based representations respectively. Section \ref{sec:implicit} further introduces implicit surface representations. Sections \ref{sec:structure} and \ref{sec:deformation} review structure-based and deformation-based description methods. We then summarize typical datasets in Section \ref{sec:datasets} and typical applications for shape analysis and reconstruction in Section \ref{sec:analysisreconstruction}, before concluding the paper in Section \ref{sec:summary}. Figure~\ref{fig:voxeltimeline} summarizes the timeline of representative deep learning methods based on  various 3D shape representations.

\section{Image-based methods}\label{sec:images}

2D images are the projections of 3D entities. Although the geometric information carried by one image is incomplete, a plausible 3D shape could be inferred from a set of images with different perspectives. The extra channel of depth in RGB-D data further enhances the capacity of image-based representations on encoding geometric cues. Benefiting from its image-like structure, the research using deep neural networks on 3D shape inferences from images started earlier than alternative representations that can depict the surface or geometry of 3D shapes explicitly. 

Socher et al.~\cite{socher2012convolutional} proposed a convolutional and recursive neural network for 3D object recognition, which copes with RGB and depth images by single convolutional layers separately and merges the features by a recursive network. Eigen et al.~\cite{eigen2014depth} first proposed to reconstruct the depth map from a single RGB image and designed a new scale invariant loss for the training stage. Gupta et al.~\cite{gupta2014learning} encoded the depth map into three channels including disparity, height and angle. Other deep learning methods based on RGB-D images are designed for 3D object detection~\cite{gupta2015aligning, song2016deep}, outperforming previous methods.

Images from different viewpoints can provide complementary cues to infer 3D objects. Thanks to the development of deep learning models in 2D fields, the learning methods based on multi-view image representation perform better in the 3D shape recognition application than those based on other 3D representations. Su et al.~\cite{su2015multi} proposed \textit{MVCNN} (Multi-View Convolutional Neural Network) for 3D object recognition. MVCNN first processes the images in different views separately by the first part of CNN, then aggregates the features extracted from different views by view-pooling layers, and finally puts the merged feature to the remaining part of CNN. Qi et al.~\cite{qi2016volumetric} propose to add multi-resolution into MVCNN for higher classification accuracy. 
%\zfl{No more works in this category?}

%2017CVPR 3D shape segmentation with projective convolutional networks

\section{Voxel-based representations}\label{voxels}

\subsection{Dense Voxel Representation}

The voxel-based representation is traditionally a dense representation, which describes 3D shape data by volumetric grids in 3D space. Each voxel in the grid records the status of occupancy (e.g., occupied or unoccupied) within a cuboid grid.

One of the earliest methods that applies deep neural networks to volumetric representations was proposed by Wu et al.~\cite{wu20153d} in 2015, which is called \textit{3D ShapeNets}. Wu et al. assigned three different states to the voxels in the volumetric representation produced by 2.5D depth maps: observed, unobserved and free. 3D ShapeNets extended the deep belief network (DBN)~\cite{hinton2006fast} from pixel data to voxel data and replaced fully connected layers in DBN with convolutional layers. The model takes the aforementioned volumetric representation as input, and outputs category labels and predicted 3D shape by iterative computations. %30*30*30
Concurrently, Maturana et al. proposed to process the volumetric representation with 3D Convolutional Neural Networks (3D CNNs)~\cite{maturana20153d} and designed \textit{VoxNet}~\cite{maturana2015voxnet} for object recognition. VoxNet defines several volumetric layers, including Input Layer, Convolutional Layers, %$C(f,d,s)$, 
Pooling Layers %$P(m)$ 
and Fully Connected Layers. %$FC(n)$. 
Although these defined layers are simple extensions of traditional 2D CNNs~\cite{krizhevsky2012imagenet} to 3D,  VoxNet is easy to implement and train and gets promising performance as the first attempt on volumetric convolutions. In addition, to ensure that VoxNet is invariant to orientation, Maturana et al. further augment the input data by rotating each shape into $n$ instances with different orientations in the training stage and adding a pooling operation after the output layer to group all the predictions from the $n$ instances in the test stage. %32*32*32

In addition to the development of deep belief networks and convolutional neural networks in shape analysis based on volumetric representation, two most successful generative models, namely auto-encoders and Generative Adversarial Networks (GANs)~\cite{goodfellow2014generative} are also extended to support this representation. Inspired by Denoising Auto-Encoders (DAEs)~\cite{vincent2008extracting, vincent2010stacked}, Sharma et al. proposed an autoencoder model \textit{VConv-DAE} for coping with voxels~\cite{sharma2016vconv}. It is one of the earliest unsupervised learning approaches in voxel-based shape analysis to our knowledge. Without object labels for training, VConv-DAE chooses mean square loss or cross entropy loss as the reconstruction loss function. %application: shape completing, denoising, recognition and interpolation 30*30*30
Girdhar et al.~\cite{Girdhar16b} also proposed \textit{TL-embedding Network}, which combine an auto-encoder for generating a voxel-based representation with a convolutional neural network for predicting the embeddings from the 2D images. %20*20*20

Choy et al. \cite{choy20163d} proposed \textit{3D-R2N2} which takes single or multiple images as input and reconstructs objects in occupancy grids. 3D-R2N2 regards input images as a sequence and designs the 3D recurrent neural network based on LSTM (Long Short-Term Memory)~\cite{hochreiter1997long} or GRU (Gated Recurrent Unit)~\cite{cho2014learning}. The architecture consists of three parts: an image encoder to extract features from 2D images, 3D-LSTM to predict hidden states as coarse representations of final 3D models, and a decoder to increase the resolution and generate target shapes. %32*32*32

Wu et al.~\cite{wu2016learning} designed a generative model called \textit{3D-GAN} that applies the Generative Adversarial Network (GAN)~\cite{goodfellow2014generative} in voxel data. 3D GAN learns to synthesize a 3D object from a sampled latent space vector $z$ with the probability distribution $P(z)$. Moreover, \cite{wu2016learning} also proposed \textit{3D-VAE-GAN} inspired by \textit{VAE-GAN}~\cite{larsen2015autoencoding} for the object reconstruction task. 3D-VAE-GAN puts the encoder before 3D-GAN for inferring the latent vector $z$ from  input 2D images and shares the decoder with the generator of 3D-GAN. %64*64*64

After the early attempts in dealing with volumetric representations by deep learning, researchers began to optimize the architecture of volumetric networks for better performance and more applications. A motivation is that the naive extension from traditional 2D domain networks often does not perform better than image-based CNNs such as  MVCNN~\cite{su2015multi}. The main challenges affecting the performance include overfitting, orientation, data sparsity and low resolution. 

Qi et al.~\cite{qi2016volumetric} proposed two new network structures aiming to improve the performance of volumetric CNNs. One introduces an extra task %which is forced to 
namely predicting class labels with subvolume space to prevent overfitting, and another utilizes elongated kernels to compress the 3D information into the 2D field in order to use 2D CNNs directly. Both of them use mlpconv  layers~\cite{lin2013network} to replace traditional convolutional layers. \cite{qi2016volumetric} also augments the input data in different orientation and elevation to encourage the network to get more local features in different poses so that the results are less influenced by orientation changes. %30*30*30
To further mitigate the orientation impact on recognition accuracy, instead of using data augmentation like \cite{maturana2015voxnet, qi2016volumetric}, \cite{sedaghat2016orientation} proposed a new model called \textit{ORION} which extends VoxNet~\cite{maturana2015voxnet} and uses a fully connected layer to predict the object class label and orientation label simultaneously. %32*32*32

%2017 3DV SEGCloud
%2018 TOG ALIGNet

\subsection{Sparse Voxel Representation (Octree)}

Voxel-based representations often lead to high computational cost because of the exponential increase of computations from pixels to voxels. Most of the methods cannot cope with or generate high-resolution models within reasonable time. For instance, \textit{TL-embedding Network}~\cite{Girdhar16b} was designed for $20^3$ voxel grids; \textit{3DShapeNets}~\cite{wu20153d} and \textit{VConv-DAE}~\cite{sharma2016vconv} were designed for $24^3$ voxel grids with 3 voxels padding on each direction of the voxel grids; \textit{VoxNet}~\cite{maturana2015voxnet}, \textit{3D-R2N2}~\cite{choy20163d} and \textit{ORION}~\cite{sedaghat2016orientation} were designed for $32^3$ voxel grids; \textit{3D-GAN} was designed for generating $64^3$ occupancy grids as 3D shape representation. As the voxel resolution increases, the occupied grids become sparser in the whole 3D space, which leads to more unnecessary computation. To address this problem, Li et al.~\cite{li2016fpnn} designed a novel method called \textit{FPNN} to cope with the data sparsity. 

Some methods instead encode the voxel grids by a sparse, adaptive data structure, namely octree~\cite{meagher1982geometric} to reduce the dimensionality of the input data. 
H{\"a}ne et al.~\cite{hane2017hierarchical} proposed {Hierarchical Surface Prediction (HSP)} that can generate voxel grids in the form of octree from coarse to fine.  H{\"a}ne et al. observed that only the voxels near the object surface need to be predicted in a high resolution, so that the proposed HSP can avoid unnecessary calculation to ensure affordable generation of high resolution voxel grids. As introduced in \cite{hane2017hierarchical}, each node in the octree is defined as a voxel block with a fixed number ($16^3$ in the paper) of voxels in different size, and each voxel block is classified into occupied, boundary and free. The decoder of the model takes a feature vector as input, and predicts feature blocks that correspond to voxel blocks hierarchically. The HSP defines that the octree has 5 layers and each voxel blocks contains $16^3$ voxels, therefore, HSP can generate up to $256^3$ voxel grids.
Tatarchenko et al.~\cite{tatarchenko2017octree} also proposed a decoder called \textit{OGN} for generating high resolution volumetric representations. In \cite{tatarchenko2017octree}, nodes in the octree are separated into three categories, including ``empty'', ``filled'' and ``mixed''. %, which are different from \cite{hane2017hierarchical}. 
%%%What are the differences? Is it not the case that empty ~ free; filled ~ occupied; mixed ~ boundary?
The octree representing a 3D model and the feature map of the octree are stored in the form of hashing tables which are indexed by the spatial position and the octree level. In order to process the feature maps represented as hash tables, Tatarchenko et al. designed a convolutional layer named \textit{OGN-Conv}, which converts the convolutional operation into matrix multiplication.  \cite{tatarchenko2017octree} adopts the method that generates different resolution of voxel grids in each decoder layer by convolutional operations in feature maps, and then decides whether to propagate the features to the next layer by specific labels (propagating the features if ``boundary'' and skipping the feature propagation if ``mixed'').

Besides the decoder model design for synthesizing voxel grids, shape analysis methods are also designed using octrees. However, conventional octree structure~\cite{meagher1982geometric} has difficulty to be used in deep networks, so many researchers try to resolve the problem by designing new structures of octrees and special operations such as convolution, pooling and unpooling on octrees. Riegler et al.~\cite{riegler2017octnet} proposed \textit{OctNet}. The octree representation mentioned in \cite{riegler2017octnet} has a relatively regular structure than a traditional octree, which places a shallow octree in regular 3D grids. The shallow octree is constrained to have up to 3 levels and is encoded in 73 bits. Each bit determines if the corresponding cell needs to be split. Wang et al.~\cite{wang2017cnn} also proposed a convolutional neural network based on octree called \textit{O-CNN}, where the model also removes pointers like shallow octree~\cite{riegler2017octnet} and stores the octree data and structure by a series of vectors including shuffle key vectors, labels and input signals. %, which is written as $(S_l, L_l, T_l)$.
%Shuffle key\cite{wilhelms1992octrees} encodes each voxel position in 3D space at octree depth $l$ with $3l$-bit code. All the shuffle keys in the same depth are stored in an integer and sorted in ascending order. \zfl{stored in an integer or as the following statement? Is the following detailed description important? Is it the basis for further discussion?} Shuffle keys in each depth are stored in a vector named $S_l$. Labels are added in order to find the parent-child relationship between adjacent layers, named $L_l$. Input signals are the average normal vector in the 3D shape surface. And the feature map vector computed by CNN in $l$-th depth is named $T_l$. Therefore, the property of depth $l$ in octree is the conmbination of shuffle keys, labels and feature map vector, . 

In addition to representing voxels, octree structure can also be utilized to represent 3D shape surfaces with planar patches. Wang et al.~\cite{wang2018adaptive} proposed \textit{Adaptive O-CNN}, where they defined another form of octree named patch-guided adaptive octree, which divides a 3D shape surface into a set of planar patches restricted by bounding boxes corresponding to octants. They also provided an encoder and a decoder for the octree defined by this paper.

\section{Surface-based representations}\label{sec:surfaces}

\subsection{Point-based Representation}

The typical point-based representation is also referred to as point clouds or point sets. They can be raw data generated by 3D scanning devices. Because of its unordered and irregular structure, this kind of representation is relatively difficult to cope with by traditional deep learning methods. Therefore, most researchers avoided to  use point clouds in a direct way at the early stage of the deep learning-based geometry research.
One of the first models to generate point clouds by deep learning came out in 2017~\cite{fan2017point}. They designed a neural network to learn a point sampler based on 3D shape point distribution. The network takes a single image and a random vector as input, and outputs an $N \times 3$ matrix representing the predicted point sets ($x$, $y$, $z$ coordinates for $N$ points). In addition, \cite{fan2017point} proposed to use \textit{Chamfer Distance (CD)} and \textit{Earth Mover's Distance (EMD)}~\cite{rubner2000earth} as the loss function to train the networks.

\begin{figure*}[!htbp]
    \centering
    \includegraphics[scale = 0.58]{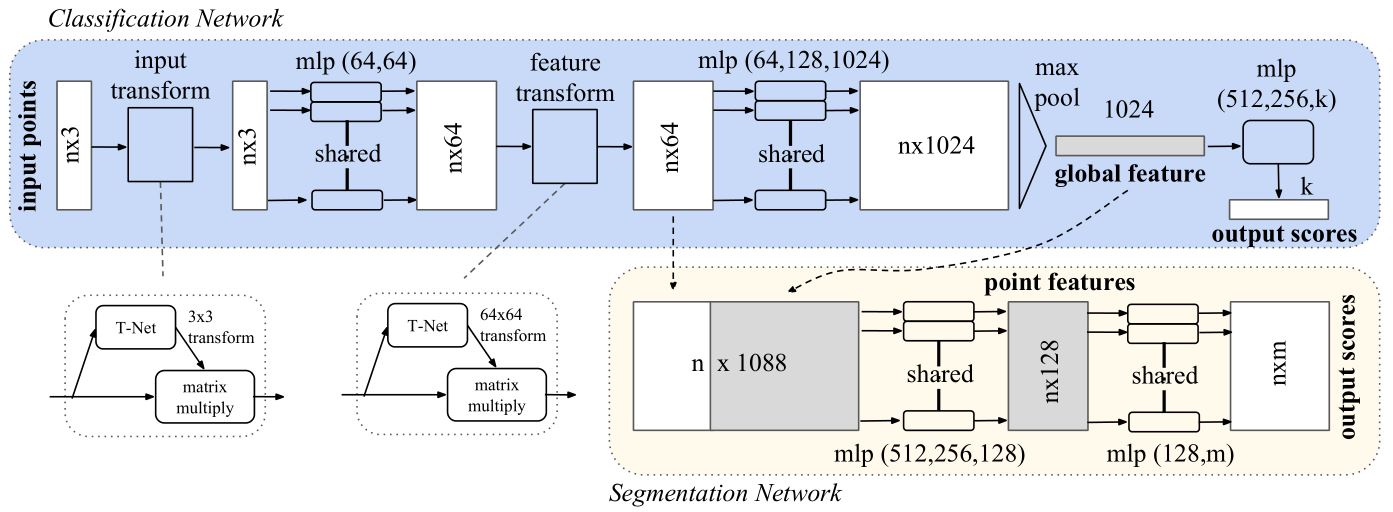}
    \caption{The pipeline of \textit{PointNet} Ref.~\cite{qi2017pointnet}, \copyright IEEE 2017.}
    \label{fig:pointnet}
\end{figure*}

\textbf{PointNet}. At almost the same time, Qi et al.~\cite{qi2017pointnet} proposed \textit{PointNet} for shape analysis, which was the first successful deep network architecture that directly processes point clouds without unnecessary rendering. The pipeline of PointNet is illustrated in Figure~\ref{fig:pointnet}.
On account of three properties of point sets mentioned in \cite{qi2017pointnet}, PointNet designed three components in their network, including using max-pooling layers as symmetry functions for dealing with the unordered property, concatenating global and local features together for point interaction, and jointly aligning the network for transformation invariance. Based on PointNet, Qi et al. further improved this model and proposed \textit{PointNet++}~\cite{qi2017pointnet++}, in order to resolve the problem that PointNet cannot capture and deal with local features induced by metric well. Compared with PointNet, PointNet++ introduces a hierarchical structure, so that the model can capture features in different scales, which improves the capability of extracting 3D shape features. As PointNet and PointNet++ show state-of-the-art performance in shape classification and semantic segmentation, more and more deep learning models were proposed based on point-based representations.

\textbf{Convolutional Neural Networks for Point Clouds}. Some research works focus on applying CNNs to the irregular and unordered form of point clouds for analysis. Li et al.~\cite{li2018pointcnn} proposed \textit{PointCNN} for point clouds and designed the $\mathcal{X}$-transformation to weight and permute the input point features, which guarantees the equivariance in different point orders. Each feature matrix needs to be multiplied by the $\mathcal{X}$-transformation matrix before passing through the convolutional operator. This process is named $\mathcal{X}$-Conv operator, which is the key of \textit{PointCNN}. Wang et al.~\cite{wang2019dgcnn}  proposed \textit{DGCNN}, a dynamic graph CNN architecture for point cloud classification and segmentation. Instead of processing point features like PointNet~\cite{qi2017pointnet}, \textit{DGCNN} first connects neighboring points in spatial or semantic space to generate a graph, and then captures the local geometry features by applying the EdgeConv operator on it. Moreover, different from other graph CNNs which process the fixed input graph, \textit{DGCNN} changes the graph to obtain new nearest neighbors in the feature space in different layers, which is beneficial to get larger and sparser receptive fields.

\textbf{Other Point Cloud Processing Techniques using Neural Networks}.  Klokov et al.~\cite{klokov2017escape} proposed \textit{Kd-Network} to process point clouds based on the form of kd-trees. Yang et al.~\cite{yang2018foldingnet} proposed \textit{FoldingNet}, an end-to-end auto-encoder for further compressing a point-based representation with unsupervised learning. Because point clouds can be transformed into 2D grids by folding operations, FoldingNet integrates folding operations in their encoder-decoder to recover input 3D shapes. Mehr et al.~\cite{mehr2019disconet} further proposed \textit{DiscoNet} for 3D model editing by combining multiple autoencoders which are trained for different types of 3D shapes specifically. The autoencoders use pre-learned mean geometry of training 3D shapes as their templates. Meng et al.~\cite{meng2019vv} proposed \textit{VV-Net} (Voxel VAE Net) for point segmentation, which represents a point cloud by a structured voxel representation. In \textit{VV-Net}, instead of containing a boolean value to represent occupancy status of each voxel as a normal volumetric representation, it uses a latent code computed by an RBF-VAE, a variational autoencoder based on a radial basis function (RBF) interpolation of points to describe point distribution within a voxel. This representation is used to extract intrinsic symmetry of point clouds by a group equivariant CNN, and the output is combined with PointNet~\cite{qi2017pointnet} for better segmentation performance.
%2018ICML "Learning Representations and Generative Models for 3D Point Clouds"

Although the point-based representation can be more easily obtained by 3D scanners than other 3D representations, this raw form of 3D shapes is often unsuitable for 3D shape analysis, due to noise and data sparsity. Therefore, compared with other representations, it is essential for the point-based representation to incorporate an upsampling module to obtain fine-grained point clouds, such as \textit{PU-NET}~\cite{yu2018pu}, \textit{MPU}~\cite{yifan2019patch}, \textit{PU-GAN}~\cite{li2019pu}, etc. 
Additionally, point cloud registration is also an essential preprocessing step, e.g. to fuse points from multiple scans, which aims to calculate rigid transformation parameters to align the point clouds. Wang et al.~\cite{wang2019dcp} proposed \textit{Deep Closest Point (DCP)}, which extends traditional Iteractive Closest Point  (\textit{ICP}) method~\cite{besl1992method} and uses a deep learning method to obtain the transformation parameters.
Recently, Guo et al.~\cite{guo2019deep} presented a survey focusing on deep learning models in point clouds, which provides more details in this field.

%xiaoguang han

\subsection{Mesh-based Representations}

Compared with point-based representations, mesh-based representations contain connectivity between neighboring points, so they are more suitable for describing local regions on surfaces. As a typical type of representation in non-Euclidean space, mesh-based representations can be processed by deep learning models both in spatial and spectral domains~\cite{bronstein2017geometric}. 

\textbf{Parametric representations for meshes}. Directly applying CNNs to irregular data structures like meshes is non-trivial, so there emerged a handful of approaches that map 3D shape surfaces to 2D domains such as 2D geometry images which can also be regarded as another 3D shape representation, and apply traditional 2D CNNs on them~\cite{sinha2016deep, maron2017convolutional}. Based on geometry images, Sinha et al.~\cite{sinha2017surfnet} proposed \textit{SurfNet} for shape generation using a deep residual network. Similarly, Shi et al.~\cite{shi2015deeppano}  projected 3D models into cylinder panoramic images, which are processed by CNNs. Some other methods convert mesh models into spherical signals, and design a convolutional operator in the spherical domain for shape analysis. To address high-resolution signals on 3D meshes, in particular texture information, Huang et al.~\cite{huang2019texturenet} proposed \textit{TextureNet} to extract features in this situation, where a 4-rotational symmetric (4-RoSy) field is defined to parametrize surfaces.
In the following, we will review deep learning models according to how meshes are directly treated as input, and introduce generative models working on meshes.

\textbf{Graphs}. The mesh-based representation is constructed by sets of vertices and edges, which can be seen as a graph. Some models were proposed based on the graph spectral theorem. They generalize CNNs on  graphs~\cite{bruna2013spectral,henaff2015deep, defferrard2016convolutional, kipf2016semi, atwood2016diffusion} by eigen-decomposition of Laplacian matrices, which is able to generalize convolutional operators to the spectral domain of graphs. Verma et al.~\cite{verma2018feastnet} proposed another graph-based CNN named \textit{FeaStNet}, which computes the receptive fields of convolution operator dynamically. Specifically, FeaStNet determines the assignment of the neighbor vertices by using features obtained in networks. Hanocka et al.~\cite{hanocka2019meshcnn} also designed operators of convolution, pooling and unpooling for triangle meshes, and proposed \textit{MeshCNN}. Different from other graph-based methods, MeshCNN focuses on processing the features stored in edges, and proposes a convolution operator that is applied to the edges with a fixed number of neighbors and a pooling operator based on edge collapse. MeshCNN extracts 3D shape features with respect to specific tasks, and the network learns to preserve the important features and ignore the unimportant ones. 

\textbf{2-Manifolds.} The mesh-based representation can be viewed as the discretization of 2-manifolds. Several works are designed in 2-manifolds with a series of refined CNN operators to adapt to this non-Euclidean space. These methods define their own local patches and kernel functions for generalizing CNN models. Masci et al.~\cite{masci2015geodesic} proposed \textit{Geodesic Convolutional Neural Networks (GCNNs)} for manifolds, which extract and discretize local geodesic patches and apply convolutional filters on these patches in polar coordinates. The convolution operator is designed in the spatial domain and their Geodesic CNN is quite similar to conventional CNNs applied in Euclidean space. \textit{Localized Spectral CNNs}~\cite{boscaini2015learning} proposed by Boscaini et al. apply \textit{Windowed Fourier transform} to non-Euclidean space. \textit{Anisotropic Convolutional Neural Networks (ACNNs)}~\cite{boscaini2016learning} further designed an anisotropic heat kernel to replace the isotropic patch operator in GCNN~\cite{masci2015geodesic}, which gives another solution to avoid ambiguity. Xu et al.~\cite{xu2017directionally} proposed \textit{Directionally Convolutional Networks (DCNs)}, which defined local patches based on faces of the mesh representation. In this work, researchers also designed a two-stream network for 3D shape segmentation, which takes local face normals and the global face distance histogram as input for training. Moti et al.~\cite{monti2017geometric} proposed \textit{MoNet} to replace the weight functions in \cite{masci2015geodesic, boscaini2016learning} with Gaussian kernels with learnable parameters. Fey et al.~\cite{fey2018splinecnn} proposed \textit{SplineCNN} which designed a convolutional operator based on B-splines. Pan et al.~\cite{pan2018convolutional} designed a surface CNN for 3D irregular surface to preserve the standard CNN property of translation equivariance by using parallel translation frames and group convolutional operations. Qiao et al.~\cite{qiao2019laplaciannet} proposed \textit{Laplacian Pooling Network (LaplacianNet)} for 3D mesh analysis. The \textit{LaplacianNet} considers both spectral and spatial information of the mesh, and contains 3 parts: preprocessing features as the network input, Mesh Pooling Blocks to split surface and cluster patches for feature extraction, and the Correlation Network to aggregate global information.

\begin{figure*}
    \centering
    \includegraphics[scale=0.46]{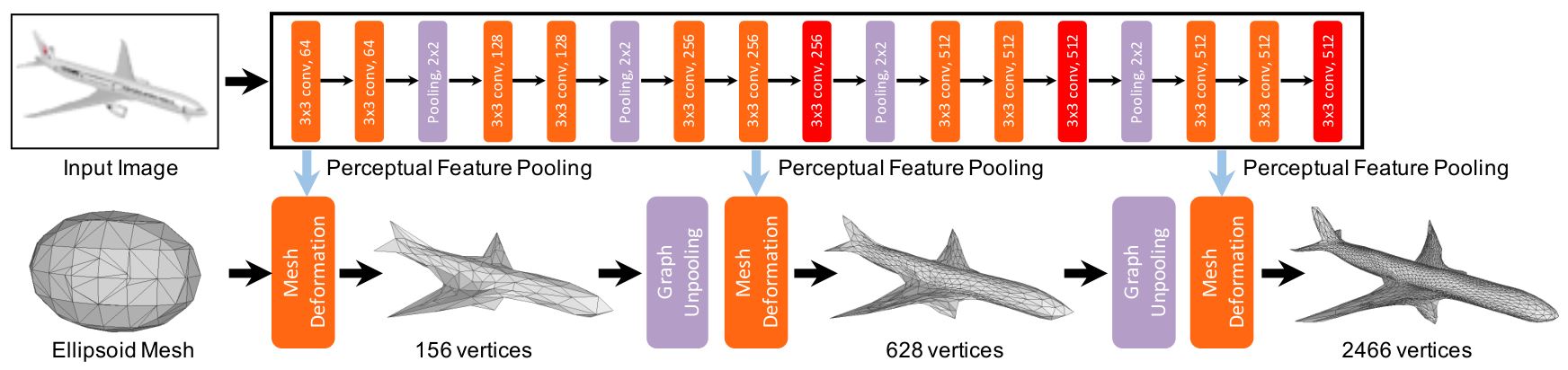}
    \caption{The pipeline of \textit{Pixel2Mesh} Ref.\cite{wang2018pixel2mesh} \copyright Springer 2018.}
    \label{fig:Pixel2Mesh}
\end{figure*}

\textbf{Generative Models.} There are also many generative models for the mesh-based representation. %Sinha et al.\cite{sinha2017surfnet}
Wang et al.~\cite{wang2018pixel2mesh} proposed \textit{Pixel2Mesh} for reconstructing 3D shapes from single images, which generates the target triangular mesh by deforming an ellipsoid template. As shown in Figure~\ref{fig:Pixel2Mesh}, the Pixel2Mesh network is implemented based on \textit{Graph-based Convolutional Networks (GCNs)}~\cite{bronstein2017geometric} and generates the target mesh from coarse to fine by an unpooling operation. Wen et al.~\cite{wen2019pixel2mesh++} advanced Pixel2Mesh and proposed \textit{Pixel2Mesh++}, which extends single image 3D shape reconstruction to 3D shape reconstruction from  multi-view images. To achieve this, Pixel2Mesh++ introduces a \textit{Multi-view Deformation Network (MDN)} to the original \textit{Pixel2Mesh}, and the \textit{MDN} incorporates the cross-view information into the process of  mesh generation. 
Groueix et al.~\cite{groueix2018atlasnet} proposed \textit{AtlasNet}, which generates 3D surfaces by multiple patches. AtlasNet learns to convert 2D square patches into 2-manifolds to cover the surface of 3D shapes by MLP (Multi-Layer Perceptron). Ben-Hamu et al.~\cite{ben2018multi} proposed a multi-chart generative model for 3D shape generation. The method uses a multi-chart structure as input and builds the network architecture based on standard image GAN~\cite{goodfellow2014generative}. The transformation between 3D surface and multi-chart structure is based on ~\cite{maron2017convolutional}. 
However, the methods based on deforming a template mesh into the target shape cannot express complex topology of some 3D shapes. Pan et al.~\cite{pan2019deep} proposed a new single-view reconstruction method, which combines a deformation network and a topology modification network to model meshes with complex topology. In the topology modification network, the faces with high distortion are removed.
Tang et al.~\cite{tang2019skeleton} proposed to generate complex topology meshes by a skeleton-bridged  learning method, because skeleton can well preserve topology information.
Instead of generating triangular meshes, Nash et al.~\cite{nash2020polygen} proposed \textit{PolyGen} to generate the polygon mesh representation. Inspired by neural autoregressive models in other fields like natural language processing, researchers regard mesh generation as a sequence, and design a transformer-based network~\cite{vaswani2017attention}, including a vertex model and a face model. The vertex model generates a sequence of vertex positions and the face model generates variable-length vertex sequences conditioned on input vertices.

\section{Implicit representations}\label{sec:implicit}

In addition to explicit representations such as point clouds and meshes, implicit fields have been in greater popularity in recent studies. A major reason is that the implicit representation is not limited by fixed topology and resolution. There are an increasing number of deep models, which define their own implicit representations and building on them further propose various methods for shape analysis and generation.

The \textbf{Occupancy/Indicator Function} is one of the forms to represent 3D shapes implicitly. \textit{Occupancy Network} was proposed by Mescheder et al.~\cite{mescheder2019occupancy} to learn a continuous occupancy function as a new representation of 3D shapes by neural networks. The occupancy function reflects the 3D point status with respect to the 3D shape surface, where 1 means inside the surface and 0 otherwise. Researchers regarded this problem as a binary classification task and designed an occupancy network which inputs 3D point position and 3D shape observation and outputs the probability of occupancy. The generated implicit field is then processed by a Multi-resolution IsoSurface Extraction method \textit{MISE} and marching cubes algorithm~\cite{lorensen1987marching} to obtain meshes. Moreover, researchers introduce  encoder networks to obtain latent embeddings. Similarly, Chen et al.~\cite{chen2019learning} designed \textit{IM-NET} as a decoder for learning generative models, which also takes an implicit function in the form of an indicator function.

\textbf{Signed Distance Functions} (\textit{SDFs}) are also a form of implicit representation. Signed distance functions map a 3D point to a real value instead of a probability, which indicates the spatial relation and distance to the 3D surface. Denote $SDF(x)$ as the signed distance value of a given 3D point $x \in \mathbb{R}^3$. Then $SDF(x) > 0$ if point $x$ is outside the 3D shape surface, $SDF(x) < 0$ if point $x$ is inside the surface, and $SDF(x) = 0$ means point $x$ is on the surface. The absolute value of $SDF(x)$ refers to the distance between point $x$ and the surface. Park et al.~\cite{park2019deepsdf} proposed \textit{DeepSDF} and introduced an \textit{auto-decoder-based DeepSDF} as a new 3D shape representation. Wang et al.~\cite{xu2019disn} also proposed \textit{Deep Implicit Surface Networks (DISNs)} for single-view 3D reconstruction based on SDFs. Thanks to the advantages of SDF, DISN was the first to reconstruct 3D shapes with flexible topology and thin structure in the single-view reconstruction task, which is difficult for other 3D representations.

\textbf{Function Sets}. The occupancy functions and signed distance functions represent the 3D shape surface by a single function learned by a deep neural network. Genova et al.~\cite{genova2019learning, genova2019deep} proposed to represent the whole 3D shape by combining a set of shape elements. In \cite{genova2019learning}, researchers proposed \textit{Structured Implicit Functions (SIFs)} where each element is represented by a \textit{scaled axis-aligned anisotropic 3D Gaussian}, and the sum of these shape elements represents the whole 3D shape. The parameters of Gaussians are learned by the CNN. \cite{genova2019deep} improved the SIF and proposed \textit{Deep Structured Implicit Functions (DSIFs)} which added deep neural networks as \textit{Deep Implicit Functions (DIFs)} to provide local geometry details. To summarize, \textit{DSIF} exploits \textit{SIF} to depict coarse information of each shape element, and applies \textit{DIF} for local shape details.

%%%YKL I think saying the method does not need 3D ground truth is not correct, but instead it does not use 3D supervision.
\textbf{Approach without 3D supervision}. The above implicit representation models need to sample 3D points in the 3D shape bounding box as ground truth and train the model supervised with 3D information. But 3D ground truth may not be easy to access in some situations. Liu et al.~\cite{liu2019learning} proposed a framework which learns implicit representations without explicit 3D supervision. The model uses a field probing algorithm to bridge the gap between the 3D shape and 2D images, and designs a silhouette loss to constrain 3D shape outline and geometry regularization to constrain the surface to be plausible.

\section{Structure-based representations}\label{sec:structure}

Recently, more and more researchers began to realize the importance of structure of 3D shapes and integrate structural information into deep learning models. Primitive representations are a typical type of structure-based representation which depict 3D shape structure well. A primitive representation represents the 3D shape with primitives such as oriented 3D boxes. Instead of providing a description of geometry details, the primitive representation concentrates more on the overall structure of 3D shapes. It represents 3D shape structure as several primitives with a compact parameter set. More importantly, obtaining a primitive representation encourages to generate more detailed and plausible 3D shapes. 

\textbf{Linearly Organized}. Observing that humans often regard 3D shapes as a collection of parts, Zou et al.~\cite{zou20173d} proposed \textit{3D-PRNN}, which applies LSTM in a primitive generator, so that 3D-PRNN can generate primitives sequentially. The generated primitive representations show great efficiency in depicting simple and regular 3D shapes. Wu et al.~\cite{wu2019pq} further proposed an RCNN-based method called \textit{PQ-NET} which also regards 3D shape parts as a sequence.  The difference is that PQ-NET encodes geometry features in the network. 
Gao et al.~\cite{gao2019sdm} proposed a deep generative model named \textit{SDM-NET} (Structured Deformable Mesh-Net). They designed a two-level VAE, containing a PartVAE for part geometry and a SP-VAE (Structured Parts VAE) for both structure and geometry features. In \cite{gao2019sdm}, each shape part is encoded in a well designed form, which records both the structure information (symmetry, supporting and supported) and geometry features.

\textbf{Hierarchically Organized}. Li et al.~\cite{li2017grass} proposed \textit{GRASS} (Generative Recursive Autoencoders for Shape Structures), which is one of the first attempts to encode the 3D shape structure by a neural network.
They describe the shape structure by a hierarchical binary tree, in which the child nodes are merged into the parent node by either adjacency or symmetry relations. Leaves in this structure tree represent the oriented bounding boxes (OBBs) and geometry features for each part, and intermediate nodes represent both the geometry feature of child nodes and the relations between child nodes. Inspired by recursive neural networks (RvNNs)~\cite{socher2011parsing, socher2012convolutional}, GRASS also recursively merges the codes representing the OBBs into a root code which depicts the whole shape structure. The architecture of GRASS can be divided into three parts: (1) an RvNN autoencoder for encoding a 3D shape into a fixed length code, (2) a GAN for learning the distribution of root codes and generating plausible structures, (3) another autoencoder for synthesizing geometry of each part which is inspired by \cite{Girdhar16b}. Furthermore, to synthesize fine-grained geometry in voxel grids, \textit{Structure-aware recursive feature (SARF)} is proposed, which contains both the geometry features of each part and global and local OBB layout. 

However, the GRASS~\cite{li2017grass} uses a binary tree to organize the part structure, which leads to ambiguity. Therefore, binary trees are not suitable for large scale datasets. To address the problem, Mo et al.~\cite{mo2019structurenet} proposed \textit{StructureNet} which organized the hierarchical structure in the form of graphs.

The \textit{BSP-Net} (Binary Space Partitioning-Net) proposed by Chen et al.~\cite{chen2019bsp} is the first method to depict sharp geometry features, which constructs a 3D shape by convexes organized by a BSP-tree. The Binary Space Partitioning (BSP) tree defined in \cite{chen2019bsp} is used to represent 3D shapes by collections of convexes, which includes three layers, namely hyperplane extraction, hyerplane grouping and shape assembly. The convexes can also be seen as a new form of primitives which can represent geometry details of 3D shapes rather than general structures.

\textbf{Structure and Geometry}. Researchers try to encode the 3D shape structure and geometry features separately~\cite{li2017grass} or jointly~\cite{wu2019sagnet}.
Wang et al.~\cite{wang2018global} proposed \textit{Global-to-Local (G2L)} generative model to generate man-made 3D shapes from coarse to fine. To address the problem that GANs cannot generate geometry details well~\cite{wu2016learning}, \textit{G2L} first applies a GAN to generate coarse voxel grids with semantic labels that represent shape structure at the global level, and then puts the voxels separated by semantic labels into an autoencoder called \textit{Part Refiner (PR)} to optimize part geometry details part by part at the local level. 
Wu et al.~\cite{wu2019sagnet} proposed \textit{SAGNet} for detailed 3D shape generation, which encodes the structure and geometry jointly by a GRU~\cite{cho2014learning} architecture in order to find intra-relation between them. The SAGNet shows better performance in tenon-mortise joints than other structure-based learning methods.

%primitive representation generation with unsupervised learning
%compact, consistent
%[Tulsiani et al. 2017] unsupervised learning. without hierarchical conception
%2019 SIGGRAPH
%volume coverage loss, surface coverage loss, mutex loss, hierarchical loss
%mask selection: tree comlete loss, mask sparsity loss, approximation loss
%semi-supervise learning

\section{Deformation-based representations}\label{sec:deformation}

Deformable 3D models play an important role in computer animation. However, most of the methods mentioned above mainly focus on rigid 3D models, while paying less attention to the deformation of non-rigid models. Compared with other representations, deformation-based representations parameterize the deformation information and have better performance when used to cope with non-rigid 3D shapes, such as articulated models.

\begin{figure}[!htb]
    \centering
    \includegraphics[scale=0.5]{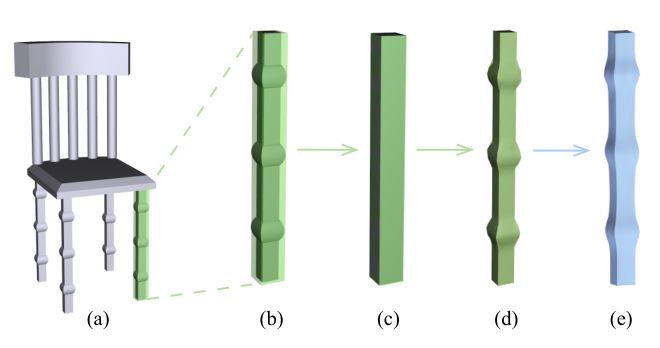}
    \caption{An example of representing a chair leg by deforming bounding box in \textit{SDM-NET}. (a)a chair with one of its leg parts highlighted, (b)the highlighted part in (a) and the overlaid bounding box, (c)the bounding box used as the template, (d)deformed bounding box, (e)recovered shape. Ref.\cite{gao2019sdm} \copyright ACM 2019}
    \label{fig:sdmnet}
\end{figure}

\textbf{Mesh-based Deformation Description}. A mesh can be seen as a graph, which is convenient when manipulating the vertex positions while maintaining the connectivity between vertices. Therefore, a great number of methods choose meshes to represent deformable 3D shapes. Based on this property, some mesh-based generation methods generate target shapes by deforming a mesh template~\cite{wang2018pixel2mesh, wen2019pixel2mesh++, pan2019deep, gao2019sdm}, and these methods can also be regarded as deformation-based methods.  The graph structure makes it easy to store deformation information as vertices features, which can be seen as deformation representations. Gao et al.~\cite{gao2016efficient} designed an efficient and rotation-invariant deformation representation called \textit{Rotation-Invariant Mesh Difference (RIMD)}, which achieves high performance in shape reconstruction, deformation and registration. Based on ~\cite{gao2016efficient}, Tan et al.~\cite{tan2018variational} proposed \textit{Mesh VAE} for deformable shape analysis and synthesis, which takes \textit{RIMD} as the feature inputs of VAE and uses fully connected layers for the encoder and decoder. Further, Gao et al.~\cite{gao2019sparse} designed an \textit{as-consistent-as-possible (ACAP) representation} to constrain the rotation angle and rotation axes between adjacent vertices in the deformable mesh which the graph convolution is easily applied. Tan et al.~\cite{tan2018mesh} proposed the \textit{SparseAE} based on the ACAP representation~\cite{gao2019sparse}, which applies graph convolutional operators~\cite{duvenaud2015convolutional} with the ACAP~\cite{gao2019sparse} to analysis the mesh deformations. Gao et al.~\cite{gao2018automatic} proposed \textit{VC-GAN} (VAE CycleGAN) for unpaired mesh deformation transfer, which is the first automatic work for mesh deformation transfer. This work takes the ACAP representation as input, and encodes the representation into latent space by a VAE, and then transfer deformations between source and target in the latent space domain with the cycle consistency and visual similarity consistency.
Gao et al. ~\cite{gao2019sdm} firstly view the geometric details shown in Fig~\ref{fig:sdmnet} as the deformations. Based on the previous techniques~\cite{gao2019sparse,gao2018automatic,tan2018variational,tan2018mesh}, the geometric details could be encoded and generated. The structure in ~\cite{gao2019sdm} is also analyzed in the stable supportable manner~\cite{huang2015support}. 
Yuan et al.\cite{yuan2019mesh} apply newly designed pooling operation based on mesh simplification and graph convolution to VAE architecture, which also takes ACAP representation as input of network. Tan et al.~\cite{tan2020realtime} use ACAP representation for simulating thin-shell deformable materials, which apply graph-based CNN to embed high-dimensional features into low-dimensional features.
In addition of considering a single deformable mesh, mesh sequences play a more important role in computer animation. And the deformation-based representation ACAP~\cite{gao2019sparse} is suitable for representing mesh sequence. Qiao et al.\cite{qiao2020synthesizing} also takes ACAP representation as input to generate mesh animation sequences by a bidirectional LSTM network.

\begin{figure*}[!htb]
    \centering
    \includegraphics[scale=0.45]{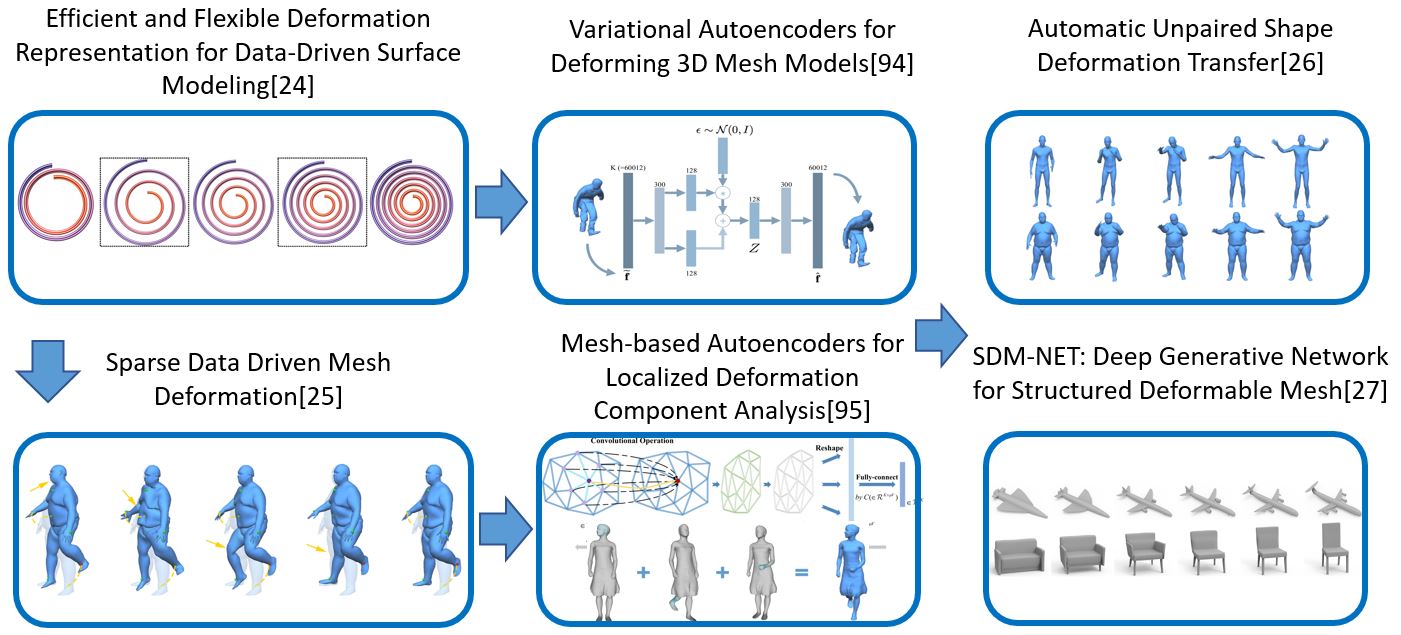}
    \caption{The research works on deformation-based shape representation of the geometrylearning group in ICT, CAS}
    \label{fig:deformation}
\end{figure*}

\textbf{Implicit surface based approaches}. With the development of implicit surface representations, Jeruzalski et al.~\cite{jeruzalski2019nasa} proposed a method to represent articulated deformable shapes by pose parameters, called \textit{Neural Articulated Shape Approximation (NASA)}. The pose parameters mentioned in \cite{jeruzalski2019nasa} record the transformation of bones defined in models. They compared three different network architectures, including unstructured model (U), piecewise rigid model (R) and piecewise deformable model (D) in the training dataset and test dataset, which opens another direction to represent deformable 3D shapes.

\section{Datasets}\label{sec:datasets}

With the development of 3D scanners, 3D models become easier to obtain, so there are more and more 3D shape datasets that have been proposed with different 3D representations. The larger datasets with more details bring more challenges for existing techniques, which further promotes the development of deep learning on different 3D representations.

The datasets can be divided into several types in different representations and different applications. Choosing the appropriate dataset benefits the performance and generalization for learning based models.

\textbf{RGB-D Images.} RGB-D image datasets can be collected by depth sensors like \textit{Microsoft Kinect}. Most of the RGB-D image datasets can be regarded as a sequence of video. The indoor scene RGB-D image dataset \textit{NYU Depth}~\cite{silberman11indoor, silberman2012indoor} was first provided for the segmentation problem, and the v1 version~\cite{silberman11indoor} collects 64 categories while the v2 version~\cite{silberman2012indoor} collects 464 categories. The \textit{KITTI}~\cite{geiger2013vision} dataset provides outdoor scene images mainly for autonomous driving, which contains 5 categories including `Road', `City', `Residential', `Campus' and `Person'. The depth map of images can be calculated by the development kit provided by the KITTI dataset. And the KITTI dataset also contains 3D objects annotations for applications such as object detection.
\textit{ScanNet}~\cite{dai2017scannet} is a large annotated RGB-D video dataset, which includes 2.5M views in 1,513 scenes with 3D camera pose, surface reconstructions and semantic segmentations.
Another dataset \textit{Human10}~\cite{Cao2018human10} is sampled from 10 human action sequences. 

%ScanNet(Dai et al. 2017) Human10(Cao et al. 2018)

\textbf{Man-made 3D Object Datasets. } The \textit{ModelNet}~\cite{wu20153d} is one of the famous CAD model datasets for 3D shape analysis, including 127,915 3D CAD Models in 662 categories. ModelNet provides two subsets named ModelNet10 and ModelNet40 respectively. ModelNet10 includes 10 categories from the whole dataset, and the 3D models in ModelNet10 are aligned manually; ModelNet40 includes 40 categories, and the 3D models are also aligned. \textit{ShapeNet}~\cite{chang2015shapenet} provides a larger scale dataset, containing more than 3 million models in more than 4K categories. ShapeNet also contains two smaller subsets: ShapeNetCore and ShapeNetSem. For various geometry applications, ShapeNet~\cite{chang2015shapenet} provides rich annotations for 3D objects in the dataset, including category labels, part labels, symmetry information, etc. 
\textit{ObjectNet3D}~\cite{xiang2016objectnet3d} is a large-scale dataset for 3D object recognition from 2D images, which includes 201,888 3D objects in 90,127 images and 44,147 different 3D shapes. The dataset is annotated with 3D pose parameters, which align 3D objects with 2D images.
\textit{SUNCG}~\cite{song2017suncg} includes full room 3D models, which is suitable for 3D scene analysis and scene completion tasks. The 3D models in \textit{SUNCG} are represented by dense voxel grids with object annotations. The whole dataset includes 49,884 valid floors with 404,058 rooms and 5,697,217 object instances.
\textit{PartNet} provides a more detailed CAD model dataset with fine-grained, hierarchical part annotations, which brings more challenges and resources for 3D object applications such as semantic segmentation, shape editing and shape generation.
3D-Future\cite{3dfuture} provides a large-scale furniture dataset, which includes 20,000+ scenes in 5,000+ rooms with 10,000+ 3D instances. Each 3D shape is of high quality with the best texture information for now.

%(Xiang et al. 2016)

\textbf{Non-Rigid Model Datasets. }
\textit{TOSCA}\cite{bronstein2008numerical} is one of the high-resolution 3D non-rigid model datasets, which contains 80 objects in 9 categories. The models are in the mesh representation, and the objects within the same category have the same resolution.
\textit{FAUST}\cite{bogo2014faust} is a dataset of 3D human body scans in 10 different people with a variety of poses and the  ground truth correspondences are also provided. Because FAUST was proposed for real-world shape registration, the scans provided in the dataset are noisy and incomplete, but the corresponding ground truth is water-tight and aligned.
\textit{AMASS}~\cite{mahmood2019amass} provides a large and varied human motion dataset, which gathers previous mocap datasets with a consistent framework and parameterization. It contains 344 subjects, 11,265 motions and more than 40 hours of recordings.

\begin{table*}[!htb]
    \centering
    \begin{tabular}{ccccccc}
        \toprule
        Source & Type & Dataset & Year & Category & Size & Description \\
        \hline
        Real-world & RGB-D Images & NYU Depth v1\cite{silberman11indoor} & 2011 & 64 & - & Indoor Scene \\
        Real-world & RGB-D Images & NYU Depth v2\cite{silberman2012indoor} & 2012 & 464 & 407024 & Indoor Scene \\
        Real-world & RGB-D Images & KITTI\cite{geiger2013vision} & 2013 & 5 & - & Outdoor Scene \\
        Real-world & RGB-D Images & ScanNet\cite{dai2017scannet} & 2017 & 1513 & 2.5M & Indoor Scene video \\
        Real-world & RGB-D Images & Human10\cite{Cao2018human10} & 2018 & 10 & 9746 & Human Action \\
        Synthetic & 3D CAD Models & ModelNet\cite{wu20153d} & 2015 & 662 & 127915 & Mesh Representation \\
        Synthetic & 3D CAD Models & ModelNet10\cite{wu20153d} & 2015 & 10 & 4899 & - \\
        Synthetic & 3D CAD Models & ModelNet40\cite{wu20153d} & 2015 & 40 & 12311 & - \\
        Synthetic & 3D CAD Models & ShpaeNet\cite{chang2015shapenet} & 2015 & 4K & 3millions & Rich Annotations \\
        Synthetic & 3D CAD Models & ShapeNetCore\cite{chang2015shapenet} & 2015 & 55 & 51300 & - \\
        Synthetic & 3D CAD Models & ShapeNetSem\cite{chang2015shapenet} & 2015 & 270 & 12000 & - \\
        Synthetic & Images and 3D Models & ObjectNet3D\cite{xiang2016objectnet3d} & 2016 & 100 & 44161 & 2D aligned with 3D \\
        Synthetic & 3D CAD Models & SUNCG\cite{song2017suncg} & 2017 & - & 49884 & Full Room Scene \\
        Synthetic & 3D CAD Models & PartNet\cite{mo2019partnet} & 2019 & 24 & 26671 & 573585 Part Instance \\
        Synthetic & 3D CAD Models & 3D-FUTURE\cite{3dfuture} & 2020 & - & 10K & Texture Information \\
        Synthetic & Non-Rigid Models & TOSCA\cite{bronstein2008numerical}&  2008 & 9 & 80 & - \\
        Real-world & Non-Rigid Models & FAUST\cite{bogo2014faust} & 2014 & 10 & 300 & Human Bodies \\
        Synthetic & Non-Rigid Models & AMASS\cite{mahmood2019amass} & 2019 & 344 & 11265 & Human Motions \\
        \bottomrule
    \end{tabular}
    \caption{The Overview of 3D Model Datasets}
    \label{tab:dataset}
\end{table*}

\section{Shape Analysis and Reconstruction}\label{sec:analysisreconstruction}

The shape representations mentioned above are fundamental for shape analysis and shape reconstruction. In this section, we summarize representative works in these two directions respectively and compare the performance of these works.

\subsection{Shape Analysis}\label{sec:analysis} 

Shape analysis methods usually extract the latent codes from different 3D shape representations by different network architectures. The latent codes are then used for specific applications like shape classification, shape retrieval, shape segmentation, etc. And different representations are usually suitable for different applications. We now review the performance of different representations in different models and discuss suitable representations for specific applications.

\textbf{Shape Classification and Retrieval} are the basic problems of shape analysis. Both of them rely on the feature vectors extracted from the analysis networks. For shape classification, the datasets ModelNet10 and ModelNet40~\cite{wu20153d} are widely used and Table \ref{tab:classification} shows the accuracy of some different methods on ModelNet10 and ModelNet40. For shape retrieval, given a 3D shape as a query, the target is to find the most similar shape(s) in the dataset to match the query. Retrieval methods usually learn to find a compact code to represent the object in a latent space, and query the closest object as the result based on Euclidean distance, Mahalanobis distance or other distance metrics. Different from the classification task, the shape retrieval task has a number of evaluation measures, including precision, recall, mAP (mean average precision), etc.

\begin{table}[!htb]
    \centering
    \begin{tabular}{cccc}
        \toprule
         \multirow{2}*{Form} & \multirow{2}*{Model} & \multicolumn{2}{c}{Accuracy(\%)} \\
         \cline{3-4}
          & & 10 & 40\\
         \hline
         Voxel & 3DShapeNet~\cite{wu20153d} & 83.54 & 77.32 \\
         Voxel & VoxNet~\cite{maturana2015voxnet} & 92 & 83 \\
         Voxel & 3D-GAN~\cite{wu2016learning} & 91.0 & 83.3 \\
         Voxel & Qi et al.~\cite{qi2016volumetric} & - & 86 \\
         Voxel & ORION~\cite{sedaghat2016orientation} & 93.8 & - \\
         Point & PointNet~\cite{qi2017pointnet} & - & 89.2 \\
         Multi-view & MVCNN~\cite{su2015multi} & - & 90.1 \\
         Point & Kd-net\cite{klokov2017escape} & 93.3 & 90.6 \\
         Multi-view & Qi et al.~\cite{qi2016volumetric} & - & 91.4 \\
         Point & PointNet++~\cite{qi2017pointnet++} & - & 91.9 \\
         Point & Point2Sequence~\cite{liu2019point2sequence} & 95.3 & 92.6 \\
         \bottomrule
    \end{tabular}
    \caption{Accuracy of shape classification on ModelNet10 and ModelNet40 datasets.}
    \label{tab:classification}
\end{table}

\textbf{Shape Segmentation} aims to discriminate the part categories of a 3D shape. This task plays an important role in understanding 3D shapes. The mean Intersection-over-Union (mIOU) is often used as the evaluation metric of shape segmentation. Most researchers choose to use the point-based representation for the segmentation task~\cite{klokov2017escape, qi2017pointnet, qi2017pointnet++, li2018pointcnn, meng2019vv}.

\textbf{Shape Symmetry Detection}. Symmetry is important geometry information in 3D shapes, and it can be further used in many other applications such as shape alignment, registration, completion, etc. Gao et al.~\cite{gao2020prs} designed the first unsupervised deep learning method named \textit{PRS-Net} (Planar Reflective Symmetry Net) to detect planar reflective symmetry of 3D shapes, which designs a new symmetry distance loss and a regularization loss. And \textit{PRS-Net} was proved to be robust in noisy and incomplete input and more efficient than traditional methods. As symmetry is largely determined by the overall shape,  \textit{PRS-Net} is based on a 3D voxel CNN and gains high performance in a low resolution.

\begin{figure*}
    \centering
    \includegraphics[scale=0.4]{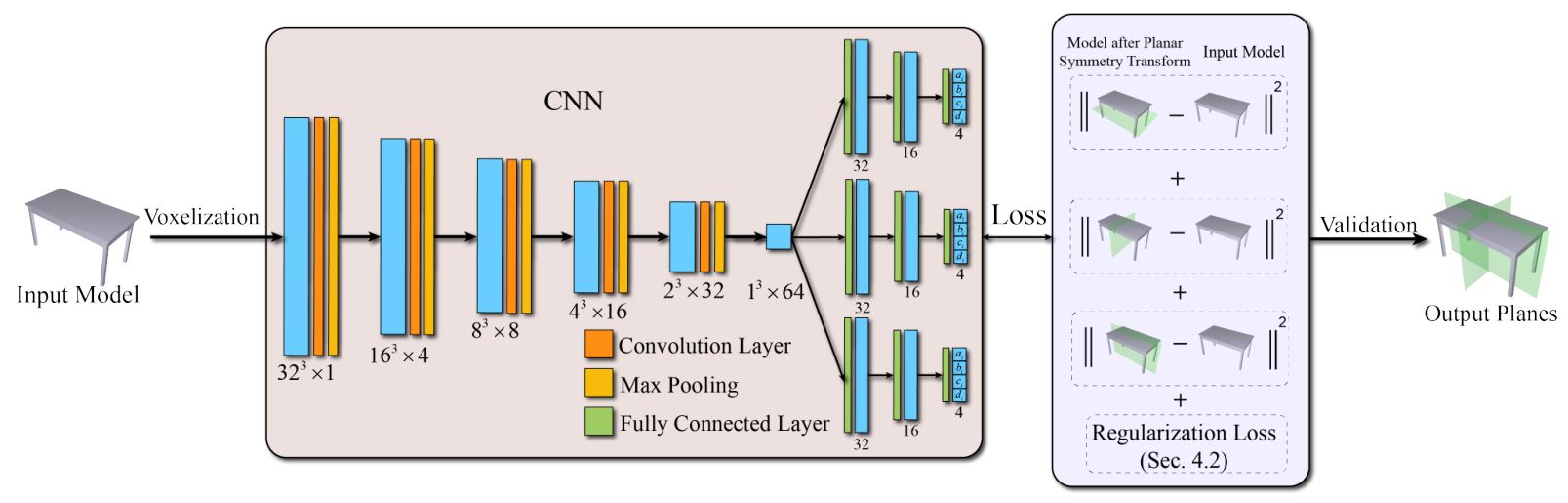}
    \caption{The pipeline of PRS-Net Ref.~\cite{gao2020prs} \copyright IEEE 2020}
    \label{fig:prsnet}
\end{figure*}

\subsection{Shape Reconstruction} 

Learning based generative models have been proposed for different representations, which is also an important field in geometry learning. The reconstruction applications include single-view shape reconstruction, shape generation, shape editing, etc. The generation methods can be summarized on the basis of representations. For voxel-based representations, learning based models try to predict the occupancy probability of each voxel in the grid. For point-based representations, learning based models either sample 3D points in the space or fold the 2D grids into target 3D objects. For mesh-based representations, most of the generation methods choose to deform a mesh template into the final mesh. In recent study, more and more methods choose to use structured representation and generate coarse-to-fine 3D shapes.

\begin{comment}
Object recognition/classification \cite{wu20153d, maturana2015voxnet, sharma2016vconv, wu2016learning, sedaghat2016orientation, riegler2017octnet, wang2017cnn, qi2017pointnet}
3D model retrieval \cite{wu20153d, wang2017cnn, li2017grass}
Scene Detection \cite{sedaghat2016orientation}
Shape generation \cite{wu20153d, sharma2016vconv, Girdhar16b, wu2016learning, hane2017hierarchical, tatarchenko2017octree, li2017grass, zou20173d, wu2019pq}
Single 2D image reconstruction \cite{Girdhar16b, wu2016learning, choy20163d, tatarchenko2017octree}
shape interpolation \cite{sharma2016vconv}
3D orientation estimation \cite{riegler2017octnet}
Semantic segmentation \cite{riegler2017octnet, wang2017cnn, qi2017pointnet}
Intrinsic correspondence \cite{masci2015geodesic}
\end{comment}

\section{Summary}\label{sec:summary}

In this survey, we review a series of deep learning methods based on different 3D object representations. We first overview different 3D representation learning models. And the tendency of the geometry learning can be summarized to be less computation and memory demanding, and more detailed and structured. Then, we introduce 3D datasets which are widely used in the research. These datasets provide rich resources and support evaluation for data-driven learning methods. Finally, we discuss 3D shape applications based on different 3D representations, including shape analysis and shape reconstruction. Different representations are usually suitable for different applications. Therefore, it is vitally important to choose suitable 3D representations for specific tasks. 

%============================================%

\CvmAck{This work was supported by National Natural Science Foundation
of China (No. 61828204 and No. 61872440), Beijing Municipal Natural Science Foundation (No. L182016), Youth Innovation Promotion Association CAS, CCF-Tencent Open Fund, Royal Society-Newton Advanced Fellowship (No. NAF{\textbackslash}R2{\textbackslash}192151) and the Royal Society (no. IES{\textbackslash}R1{\textbackslash}180126).}

\bibliographystyle{CVM}
\bibliography{ref}

\Author{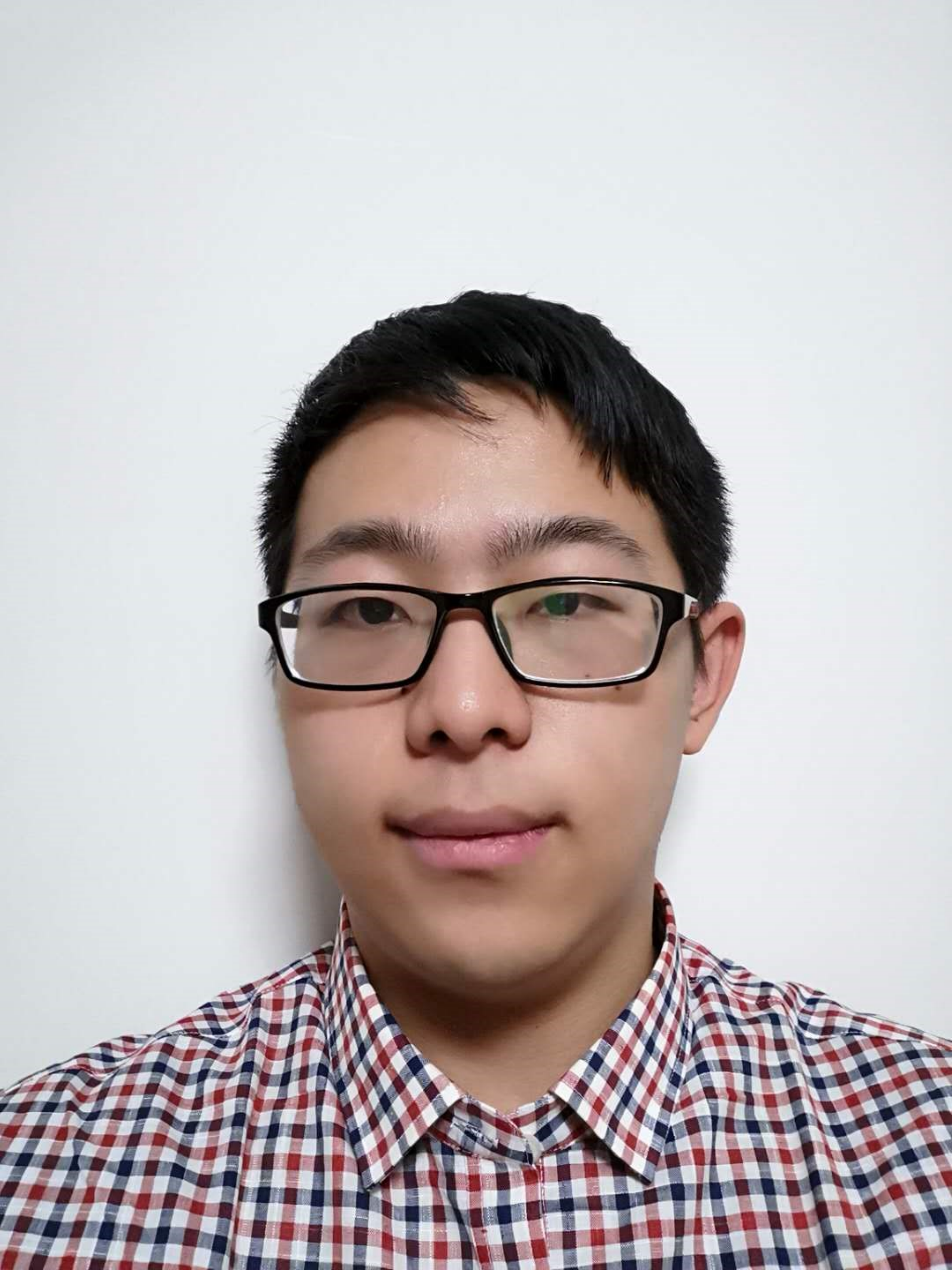}{Yun-Peng Xiao}
{received his bachelor's degree in computer science from Nankai University, He is currently a Master Student in the Institute of Computing Technology, Chinese Academy of Sciences. His research interests include computer graphics and geometric processing.}

\Author{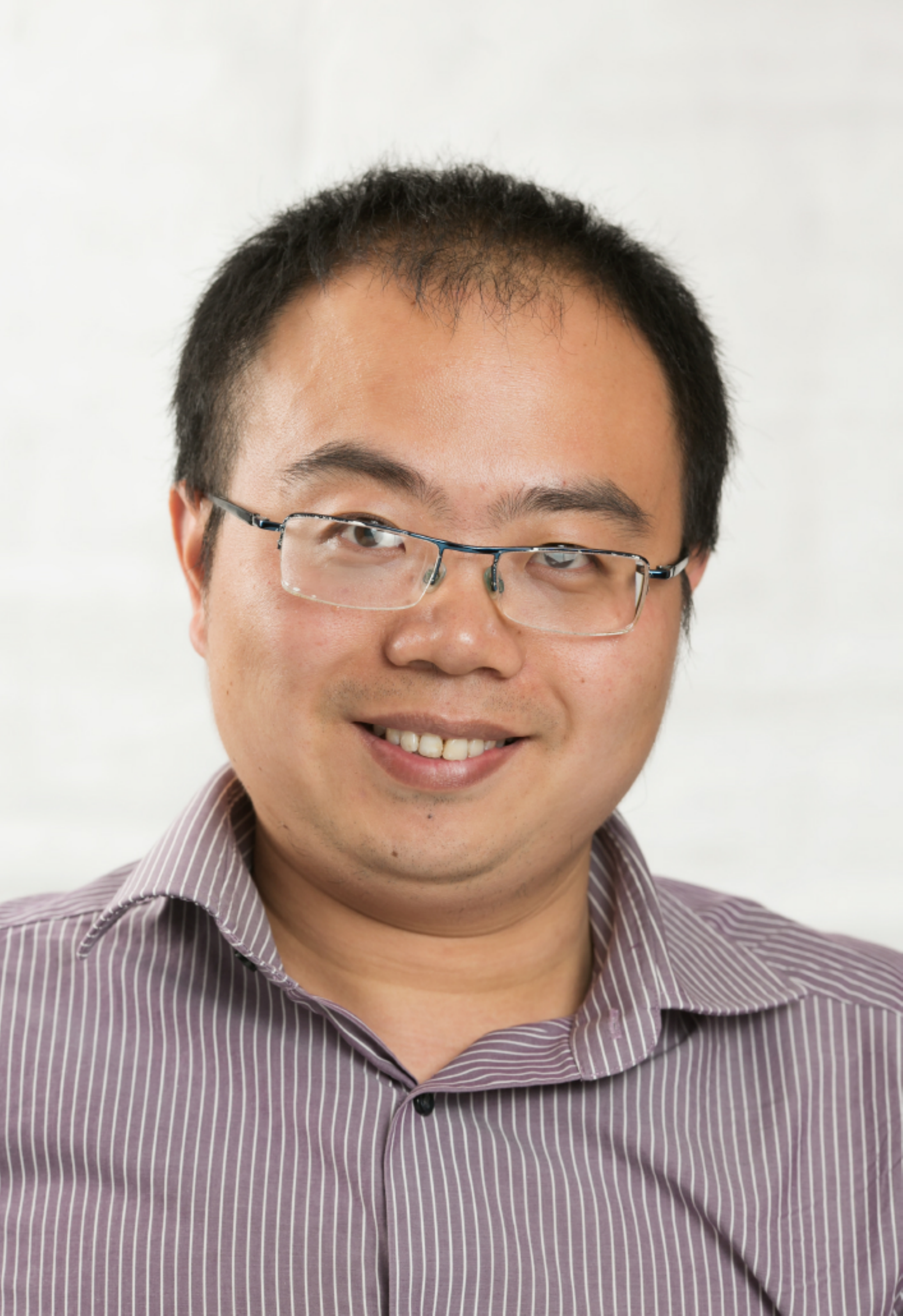}{Yu-Kun Lai}
{received his bachelor's degree and PhD degree in computer science from Tsinghua University in 2003 and 2008, respectively. He is currently a Reader in the School of Computer Science \& Informatics, Cardiff University. His research interests include computer graphics, geometry processing, image processing and computer vision. He is on the editorial boards of \emph{Computer Graphics Forum} and \emph{The Visual Computer}.}

\Author{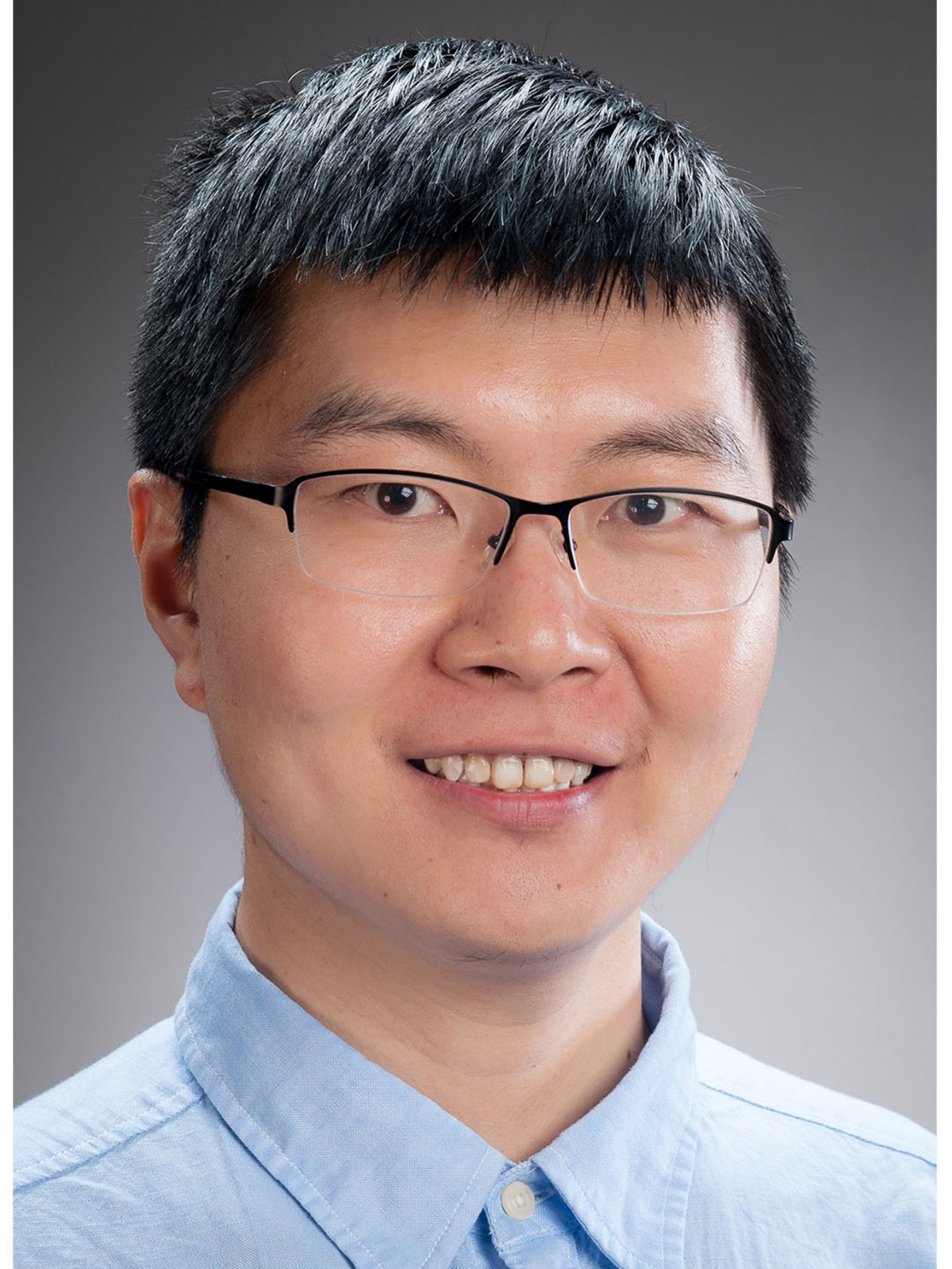}{Fang-Lue Zhang}
{is currently a Lecturer with Victoria University of Wellington, New Zealand. He received the Bachelor’s degree from Zhejiang University, Hangzhou, China, in 2009, and the Doctoral degree from Tsinghua University, Beijing, China, in 2015. His research interests include image and video editing, computer vision, and computer graphics. He is a member of IEEE and ACM. He received Victoria Early-Career Research Excellence Award in 2019.}

\Author{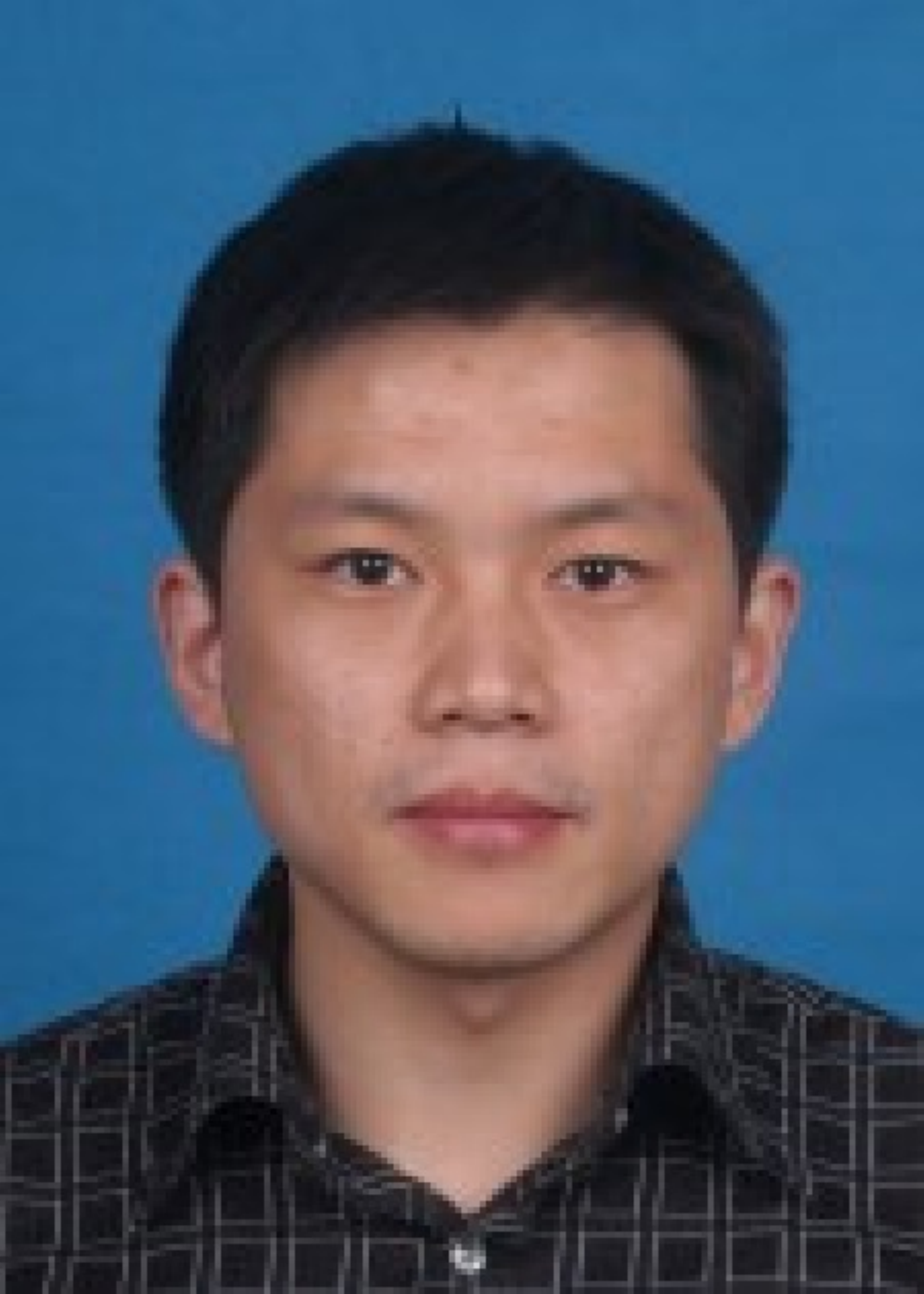}{Chunpeng Li}
{was born in 1980. He received his PhD degree in 2008 and now is an Associate Professor at the Institute of Computing Technology, Chinese Academy of Sciences. His main research interests are virtual reality, human–computer interaction, and computer graphics.}

\Author{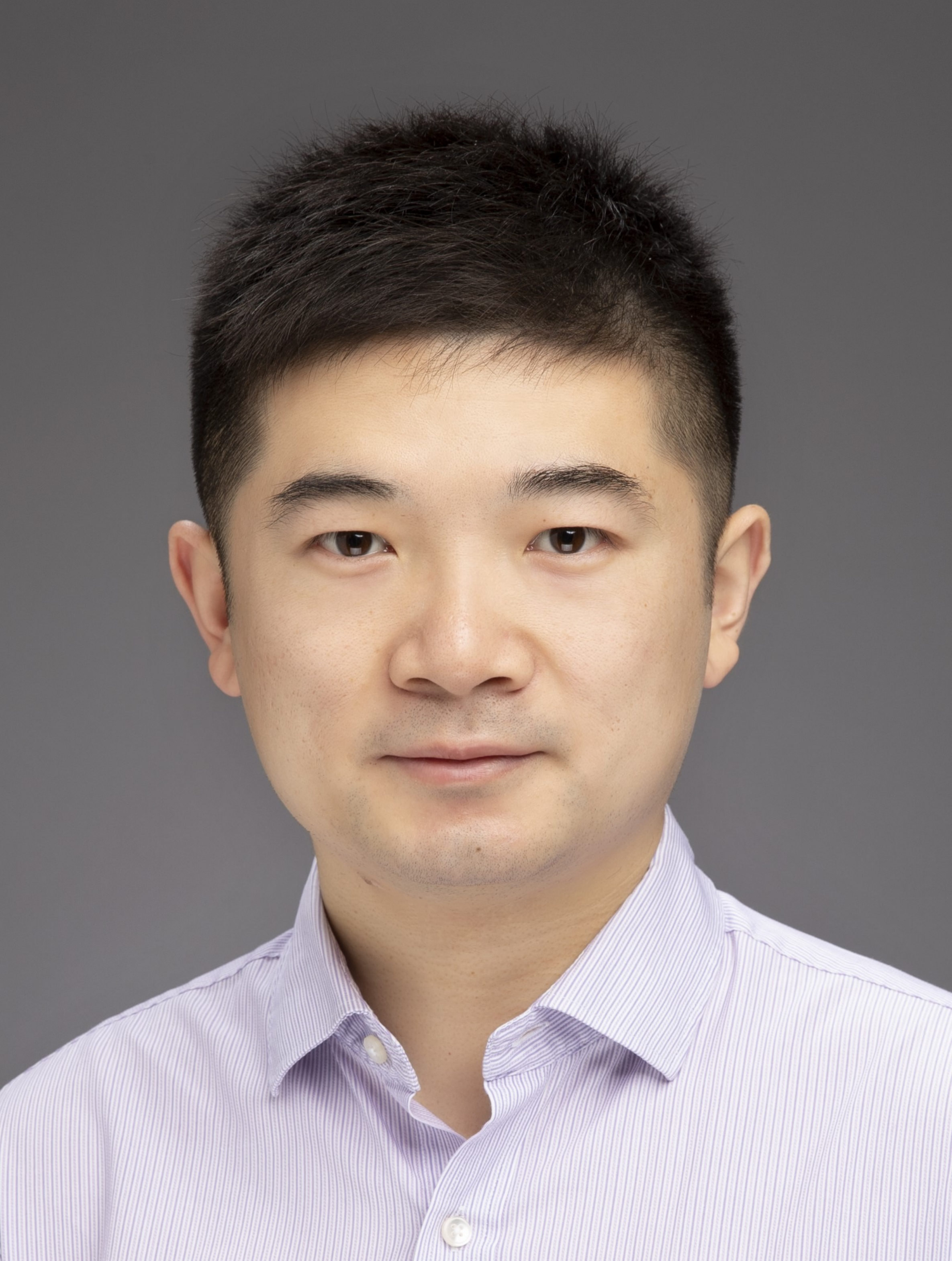}{Lin Gao}
{received a bachelor's degree in mathematics from Sichuan University and a PhD degree in computer science from Tsinghua University. He is currently an Associate Professor at the Institute of Computing Technology, Chinese Academy of Sciences. His research interests include computer graphics and geometric processing. He received the Newton Advanced Fellowship award from the Royal Society in 2019.}

\end{document}